\documentclass[nonblindrev]{informs3} 

\usepackage{balance}
\usepackage{amsmath}
\usepackage[tight,footnotesize]{subfigure}
\usepackage{graphicx}
\usepackage{bm}
\usepackage{algorithm}
\usepackage[noend]{algorithmic}
\usepackage{multirow}
\usepackage{diagbox}
\usepackage{caption}
\usepackage{amsfonts}
\usepackage{siunitx}

\OneAndAHalfSpacedXII 

\usepackage{natbib}
 \bibpunct[, ]{(}{)}{,}{a}{}{,}%
 \def\bibfont{\small}%
\TheoremsNumberedThrough     

\EquationsNumberedThrough    

\MANUSCRIPTNO{2015} 

\begin{document}


\RUNAUTHOR{Tang, Cai, Yuan and Han}

\RUNTITLE{Assortment Planning with Sponsored Products}

\TITLE{Assortment Planning with Sponsored Products}

\ARTICLEAUTHORS{%
\AUTHOR{Shaojie Tang, Shuzhang Cai}
\AFF{Naveen Jindal School of Management, The University of Texas at Dallas}
\AUTHOR{Jing Yuan}
\AFF{Department of Computer Science and Engineering, The University of North Texas}
\AUTHOR{Kai Han}
\AFF{School of Computer Science and Technology, Soochow University}
} 

\ABSTRACT{In the rapidly evolving landscape of retail, assortment planning plays a crucial role in determining the success of a business. With the rise of sponsored products and their increasing prominence in online marketplaces, retailers face new challenges in effectively managing their product assortment in the presence of sponsored products. Remarkably, previous research in assortment planning largely overlooks the existence of sponsored products and their potential impact on overall recommendation effectiveness. Instead, they commonly make the simplifying assumption that all products are either organic or non-sponsored.  This research gap underscores the necessity for a more thorough investigation of the assortment planning challenge when sponsored products are in play. We formulate the assortment planning problem in the presence of sponsored products as a combinatorial optimization task. The ultimate objective is to compute an assortment plan that optimizes expected revenue while considering the specific requirements of placing sponsored products strategically. 
}


\maketitle
\section{Introduction}

Assortment planning plays a pivotal role in the success of e-commerce platforms and recommendation systems. It involves the strategic presentation of a diverse selection of products to users, with the ultimate objective of maximizing platform revenue. By offering a tailored assortment that aligns with users' preferences and interests, platforms can enhance user experience and foster customer satisfaction.

\begin{figure*}[hptb]
\begin{center}
\hspace*{-0.25in}
\includegraphics[scale=0.6]{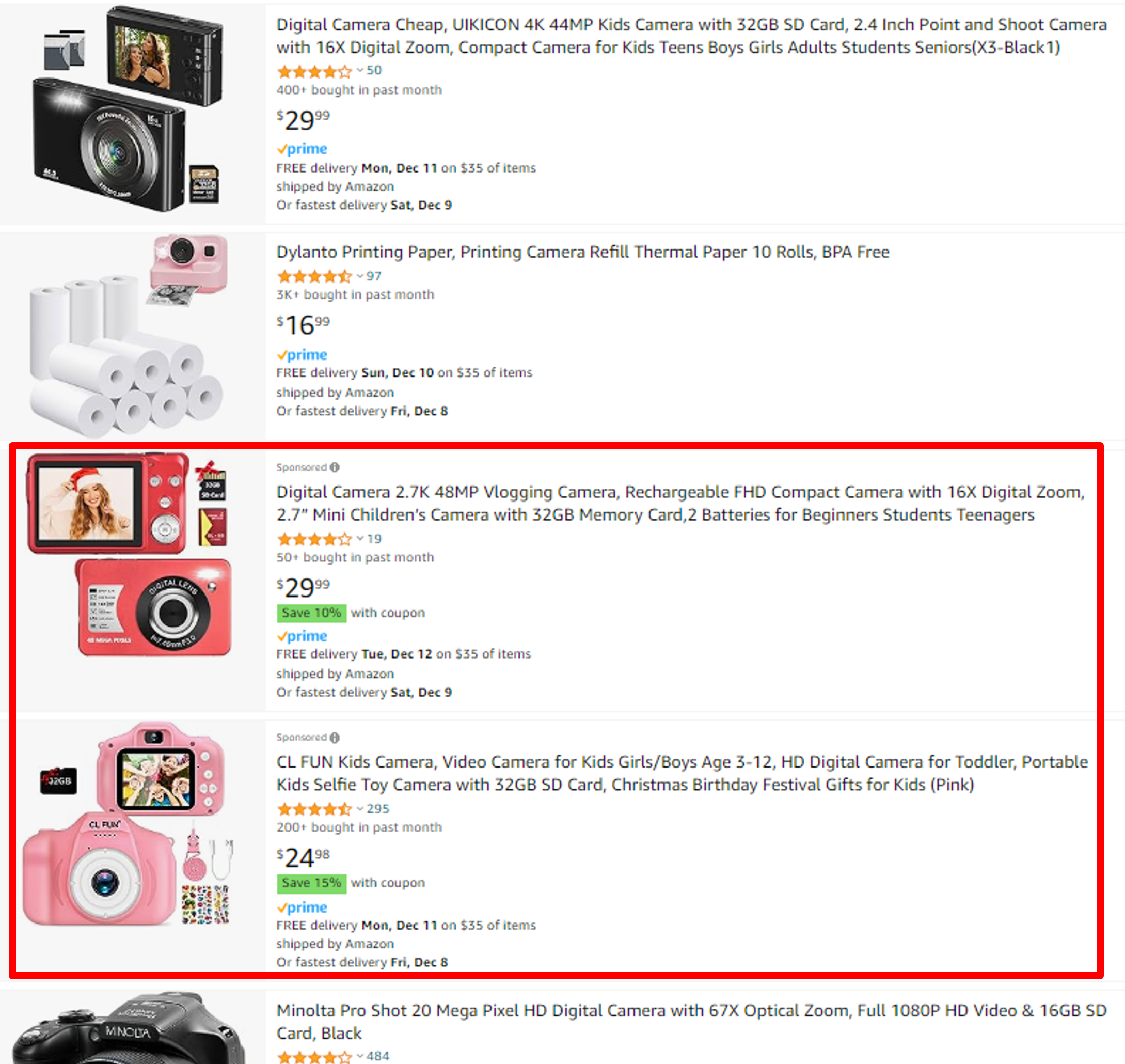}
\caption{A recommendation system displaying a list of cameras, with sponsored products highlighted in the red box.}
\label{fig:rev-cm0}
\end{center}
\end{figure*}

Prior research in assortment planning \citep{talluri2004revenue} has predominantly concentrated on organic products, which are displayed on user websites without charge to brands and can appear in arbitrary positions. Nevertheless, the e-commerce landscape has evolved in recent years with the introduction of sponsored placements. These sponsored products, frequently promoted by advertisers, have emerged as a progressively substantial source of revenue for platforms like Amazon. An example of such a recommendation system is provided in Figure \ref{fig:rev-cm0}.

However, their integration presents a nuanced challenge for both platform operators and advertisers \citep{liao2022cross,carrion2023blending,yan2020ads}.  On one hand, the primary goal of the platform remains maximizing revenue and ensuring a positive user experience. This entails providing relevant and valuable product recommendations that cater to individual user preferences. On the other hand, advertisers seek to promote specific products or brands that might not always align perfectly with the platform's overall revenue generation strategy. Striking the right balance between satisfying the platform's objectives and meeting the advertiser's goals is crucial to the success of the assortment planning process.

This dilemma  gives rise to an essential research question:

\emph{``How can businesses develop an assortment plan that incorporates the requirement of including specific sponsored products while maximizing the platform's revenue?''}

Remarkably, existing research in assortment planning has often overlooked the presence of sponsored products and their potential impact on overall recommendation performance. Although \citep{liao2022cross} is one of the few studies that consider the integration of sponsored and organic products, their problem setting differs from ours. Furthermore, their contribution lies in the development of deep learning-based heuristics without offering theoretical guarantees for their solutions. This research gap underscores the necessity for a more comprehensive investigation of the assortment planning problem when both organic and sponsored products are involved.

In this paper, we propose leveraging the Multinomial Logit (MNL) model  \citep{anderson1992discrete,mcfadden1973conditional} as the consumer's choice model, which allows us to capture user preferences effectively. We formulate the assortment planning problem in the presence of sponsored products as a combinatorial optimization task. The ultimate objective is to compute an assortment plan that optimizes expected revenue while considering the specific requirements of placing sponsored products strategically.

Through the development of a set of assortment planning algorithms with performance guarantees, our approach aims to bridge the gap between the platform's interests and advertiser goals. We seek to provide businesses with valuable insights into how they can integrate sponsored products seamlessly into their assortment plans without compromising the overall revenue-generation strategy or user experience.

\section{Preliminaries and Problem Statement}

\begin{figure*}[hptb]
\begin{center}
\hspace*{-0.25in}
\includegraphics[scale=0.5]{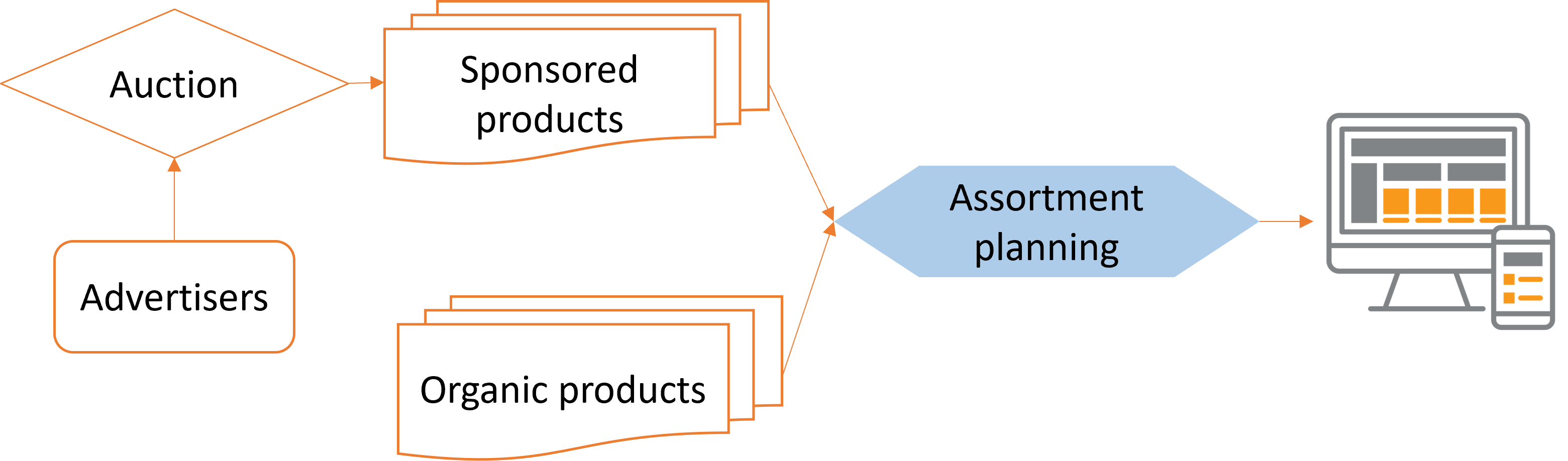}
\caption{Structure of an assortment selection system.  Our primary focus is on designing an  assortment planning module.}
\label{fig:rev-cm1}
\end{center}
\end{figure*}

Our problem involves two types of products, namely, \emph{sponsored products} and \emph{organic products}. Let $\mathcal{S}$ be the set of sponsored products that the platform decides to promote within the assortment plan; $\mathcal{O}$ be the set of organic (non-sponsored) products available for recommendation. Here $\mathcal{S}$ is chosen through some kind of auction or bidding process as illustrated in Figure \ref{fig:rev-cm1}. Each product $i\in \mathcal{O} \cup \mathcal{S}$ is associated with revenue denoted by $r_i$, indicating the revenue collected by the platform when consumers engage with the product $i$.
For organic products, this revenue can be interpreted as the commission fee collected by the platform. In contrast, for sponsored products, the revenue may include both the commission fee and the pay-per-click earnings from advertising.

\subsection{Choice model}

Given $k$ available positions for displaying products, we can assume that the intrinsic utility of product $i$ when placed in position $t$ is represented as $w(i, t)$. Now, consider a sequence of products $\pi = \{\pi_1, \cdots, \pi_k\}$, where for each product $i$ in $\pi$, we can denote the position of product $i$ in $\pi$ as $\pi^{-1}(i)$.  Hence $\pi^{-1}(\pi_t)=t$. Given an assortment $\pi$, the probability  that a consumer chooses $i \in \pi$ under the MNL model  \citep{anderson1992discrete,mcfadden1973conditional}  is
\begin{eqnarray}
\theta_{i}(\pi) = \frac{w(i, \pi^{-1}(i))}{w_0+\sum_{j\in \pi}w(j, \pi^{-1}(j)) }
\end{eqnarray}
where $w_0$ denotes the intrinsic utility of the non-purchase option. Hence, the expected revenue of displaying $\pi$ is
\begin{eqnarray}
f(\pi) = \sum_{i\in \pi}r_i\cdot \theta_i(\pi).
\end{eqnarray}

Our goal is to determine a \emph{feasible} sequence of $k$ products $\pi=\{\pi_1, \cdots, \pi_k\}$ over $\mathcal{O} \cup \mathcal{S}$ that maximizes the expected revenue. We will now explain what constitutes a feasible sequence.

\subsection{Feasible sequence} Suppose there are $k$ available positions denoted by $\mathcal{P}\cup \mathcal{R}=\{1, \cdots, k\}$, where $\mathcal{P}$ represents \emph{organic positions} designated for organic products, and $\mathcal{R}$ represents the \emph{reserved positions} for sponsored products. Without loss of generality, we assume that $|\mathcal{S}|=|\mathcal{R}|$, meaning there are sufficient reserved positions to accommodate all products that the platform decides to sponsor.

 Each sponsored product $i\in \mathcal{S}$ is associated with a set of \emph{valid} positions $\mathcal{R}_i\subseteq \mathcal{R}$.  It is a mandatory requirement for the platform to assign each sponsored product $i \in \mathcal{S}$ to one of its valid positions within $\mathcal{R}_i$. In the context of online advertising, it is common to employ an auction mechanism for positioning and prioritizing sponsored products. Advertisers compete by bidding for specific ad placements, and the outcome of these auctions determines $\mathcal{R}_i$ for each sponsored product $i\in \mathcal{S}$.  To demonstrate the generality of the proposed model, we examine its application in two real-world auction scenarios for assigning sponsored products:

Scenario 1: Social media platforms such as Facebook and Instagram offer ad placement options where advertisers can bid for specific ad positions or placements within users' feeds or on specific pages. In this case, we can interpret $\mathcal{R}_i$ as the set of positions that the advertiser $i \in \mathcal{S}$ has specified. Similarly, Amazon allows advertisers to bid on top-$m$ positions for some constant $m$. In such instances, we can define $\mathcal{R}_i$ as the set containing positions from $1$ to $m$ for all $i \in \mathcal{S}$ that bid on the top-$m$ positions.

Scenario 2: Google Ads does not provide a direct option for advertisers to bid on specific positions for sponsored placements. Instead, it determines the position of each ad based on its quality score. In this case, $\mathcal{R}_i$ is a singleton set that contains the position assigned to sponsored product $i$.

 We say a sequence $\pi$ is feasible if, for all sponsored products $i\in \mathcal{S}$, it holds that $\pi^{-1}(i)\in \mathcal{R}_i$. In other words, a sequence $\pi$ is feasible if it ensures that all sponsored products are placed in one of their valid positions.

\subsection{Problem formulation}

Recall that our goal is to determine a \emph{feasible} sequence of $k$ products that maximizes the expected revenue. The formulation can be represented as follows:

 \begin{center}
\framebox[0.4\textwidth][c]{
\enspace
\begin{minipage}[t]{0.4\textwidth}
\small
$\textbf{P.0}$
$\max_{\pi: |\pi|\leq k}  f(\pi)$ \\
subject to $\forall i\in \mathcal{S}$, $\pi^{-1}(i)\in \mathcal{R}_i$.
\end{minipage}
}
\end{center}
\vspace{0.1in}

\section{Optimal Algorithm}
\label{sec:1}

To capture the assortment planning decisions, we use $x=\{x_{it}: i\in \mathcal{O} \cup \mathcal{S}, t\in \mathcal{P}\cup \mathcal{R}\}$, where $x_{it}=1$ indicates that product $i$ is placed in position $t$, otherwise, $x_{it}=0$. Given an assortment planning $x$, its expected revenue can be calculated as follows:
\begin{eqnarray}
\frac{r_i\cdot w(i, t)\cdot x_{it}}{w_0+\sum_{i\in \mathcal{O} \cup \mathcal{S}}\sum_{t\in  \mathcal{P}\cup \mathcal{R}}w(i, t)\cdot x_{i,t} }.
\end{eqnarray}

For each sponsored position $t \in \mathcal{R}$, let $\mathcal{S}_t$ represent the set of sponsored products that have position $t$ specified as valid. That is, $\mathcal{S}_t=\{i\in \mathcal{S}\mid t\in \mathcal{R}_i\}$. Therefore, problem $\textbf{P.0}$ is reduced to
 \begin{center}
\framebox[0.4\textwidth][c]{
\enspace
\begin{minipage}[t]{0.4\textwidth}
\small
$\textbf{P.1}$
$\max_{x}  \frac{r_i\cdot w(i, t)\cdot x_{it}}{w_0+\sum_{i\in \mathcal{O} \cup \mathcal{S}}\sum_{t\in  \mathcal{P}\cup \mathcal{R}}w(i, t)\cdot x_{i,t} }$ \\
subject to \\
$\sum_{t\in  \mathcal{R}_i} x_{it}=1, \forall i\in \mathcal{S}$\\
$\sum_{i\in  \mathcal{S}_t} x_{it}=1, \forall t\in \mathcal{R}$\\
$\sum_{t\in  \mathcal{P} } x_{it}\leq1, \forall i\in \mathcal{O}$\\
$\sum_{i\in  \mathcal{O} } x_{it}\leq1, \forall t\in \mathcal{P}$\\
$x_{it}\in\{0,1\}, \forall i\in \mathcal{O} \cup \mathcal{S}, t\in \mathcal{P}\cup \mathcal{R}$
\end{minipage}
}
\end{center}
\vspace{0.1in}

The first pair of constraints, namely $\sum_{t\in \mathcal{R}_i} x_{it}=1, \forall i\in \mathcal{S}$ and $\sum_{i\in \mathcal{S}_t} x_{it}=1, \forall t\in \mathcal{R}$, guarantee that each sponsored product is allocated to one of its valid positions. The subsequent two sets of constraints, i.e., $\sum_{t\in \mathcal{P}} x_{it}\leq1, \forall i\in \mathcal{O}$ and $\sum_{i\in \mathcal{O}} x_{it}\leq 1, \forall t\in \mathcal{P}$, ensure that, at most, one organic product is assigned to each organic position.

The constraint matrix in $\textbf{P.1}$ corresponds to an assignment problem, which is well-known for being totally unimodular. It turns out solving $\textbf{P.1}$  is reduced to solving an equivalent Linear Programming (LP) problem that is guaranteed to yield an integer solution. Notably, this integer solution happens to be the optimal solution of $\textbf{P.1}$ as well \citep{davis2013assortment}.

\section{Extension: Incorporating additional constraints on organic products}

In this section, we explore an extended model of $\textbf{P.0}$ by introducing additional constraints related to the selection of organic products. Specifically, we consider a feasible family $\mathcal{I} \subseteq 2^\mathcal{O}$ of organic products. This family $\mathcal{I}$ adheres to the property of being downward closed, meaning that if a set $A \in \mathcal{I}$ and another set $B \subseteq A$, then $B \in \mathcal{I}$. A set $A$ is deemed \emph{feasible} if and only if it belongs to $\mathcal{I}$. Notably, one special case of $\mathcal{I}$ encompasses subsets of $\mathcal{O}$ that satisfy capacity (or knapsack) constraints. For example, assuming each product $i\in \mathcal{O}$ has a cost of $c_i$, and we are also given a capacity constraint $C$. In this context, the feasible family
 $\mathcal{I}$ can be defined as  $\mathcal{I}=\{A\subseteq \mathcal{O} \mid \sum_{i\in A} c_i \leq C\}$. A formal definition of this problem is listed in $\textbf{P.2}$. Here, we are abusing notation by using $\pi$ to represent both a sequence of products and the set of products contained in that sequence.

 \begin{center}
\framebox[0.4\textwidth][c]{
\enspace
\begin{minipage}[t]{0.4\textwidth}
\small
$\textbf{P.2}$
$\max_{\pi: |\pi|\leq k}  f(\pi)$ \\
subject to \\
$\forall i\in \mathcal{S}$, $\pi^{-1}(i)\in \mathcal{R}_i$.\\
$\pi\cap  \mathcal{O} \in \mathcal{I}$.
\end{minipage}
}
\end{center}
\vspace{0.1in}

Before presenting our algorithm, we first analyze the structure of the optimal solution of $\textbf{P.2}$. Let $\pi^*$ denote the optimal solution of $\textbf{P.2}$, the expected revenue of  $\pi^*$ is
\begin{eqnarray}
f(\pi^*) &&= \sum_{i\in \pi}r_i\cdot \theta_i(\pi^*) = \sum_{i\in \pi^*} \frac{r_i\cdot  w(i, {\pi^*}^{-1}(i))}{w_0+\sum_{j\in \pi^*}w(j, {\pi^*}^{-1}(j)) }\\
&&= \underbrace{\sum_{i\in \pi^*\cap \mathcal{S}} \frac{r_i\cdot  w(i, {\pi^*}^{-1}(i))}{w_0+\sum_{j\in \pi^*}w(j, {\pi^*}^{-1}(j)) } }_{\texttt{part I}}+ \underbrace{\sum_{i\in \pi^*\cap  \mathcal{O} } \frac{r_i\cdot  w(i, {\pi^*}^{-1}(i))}{w_0+\sum_{j\in \pi^*}w(j, {\pi^*}^{-1}(j)) }}_{\texttt{part II}} \label{eq:8}.
\end{eqnarray}
Intuitively, \texttt{part I} (\texttt{part II} resp.) in (\ref{eq:8}) represents the expected revenue obtained from sponsored products (organic products resp.) in  $\pi^*$.

The core idea behind our algorithm is to calculate two candidate assortments that approximate the revenue of \texttt{part I} and \texttt{part II} in (\ref{eq:8}), respectively. Subsequently, by selecting the superior solution from these two candidates, we can obtain an approximate solution for the original problem $\textbf{P.2}$.

\textbf{Candidate assortment I:} Recall that  \texttt{part I} in  (\ref{eq:8}) represents the revenue obtained from sponsored products in  $\pi^*$. Suppose we can solve the following problem optimally, we can obtain an assortment whose expected revenue is at least  \texttt{part I}.

 \begin{center}
\framebox[0.4\textwidth][c]{
\enspace
\begin{minipage}[t]{0.4\textwidth}
\small
$\textbf{P.3}$
$\max_{\pi: \pi \cap \mathcal{O}=\emptyset}  f(\pi)$ \\
subject to $\forall i\in \mathcal{S}$, $\pi^{-1}(i)\in \mathcal{R}_i$.
\end{minipage}
}
\end{center}
\vspace{0.1in}

It is easy to verify that $\textbf{P.3}$ is a special case of $\textbf{P.0}$, by setting $\mathcal{O}=\emptyset$. Hence, we can adopt our solution developed in Section \ref{sec:1} to find an optimal solution of $\textbf{P.3}$. Let $\pi^I$ denote such a solution, one can verify that  $\pi^I$ is a feasible solution of  $\textbf{P.2}$ given that $\mathcal{I}$ is defined by downward-closed constraints.
\begin{lemma}
\label{lem:2}
 Let $\pi^I$ denote the optimal solution of $\textbf{P.3}$, we have $f(\pi^I) \geq \sum_{i\in \pi^*\cap \mathcal{S}} \frac{r_i\cdot  w(i, {\pi^*}^{-1}(i))}{w_0+\sum_{j\in \pi^*}w(j, {\pi^*}^{-1}(j)) } $. I.e.,$f(\pi^I) \geq \mbox{ \texttt{part I} in  (\ref{eq:8})}$.
\end{lemma}
\emph{Proof:} Observing that employing the assortment planning  $\pi^*$ for allocating all sponsored products results in a feasible solution for $\textbf{P.3}$. This, together with the assumption that $\pi^I$ is the optimal solution of $\textbf{P.3}$, implies that \begin{eqnarray}
f(\pi^I) &&\geq \sum_{i\in \pi^*\cap \mathcal{S}} \frac{r_i\cdot  w(i, {\pi^*}^{-1}(i))}{w_0+\sum_{j\in \pi^*\cap \mathcal{S}}w(j, {\pi^*}^{-1}(j)) } \\
&&\geq \sum_{i\in \pi^*\cap \mathcal{S}} \frac{r_i\cdot  w(i, {\pi^*}^{-1}(i))}{w_0+\sum_{j\in \pi^*}w(j, {\pi^*}^{-1}(j)) }.
\end{eqnarray} $\Box$

\textbf{Candidate assortment II:} We next focus on finding another candidate assortment whose revenue is an approximation of  \texttt{part II} in (\ref{eq:8}).  The basic idea of our solution is to first compute a feasible assortment of sponsored products to minimize the total weight of the selected products. Subsequently, on top of the previously selected sponsored products, we compute the best assortment of organic products.

Step 1: We first introduce the following problem whose objective is to compute an assortment of sponsored products that has the minimum total weight.
\begin{center}
\framebox[0.4\textwidth][c]{
\enspace
\begin{minipage}[t]{0.4\textwidth}
\small
$\textbf{P.4}$
$\min_{\pi: \pi \cap \mathcal{O}=\emptyset}  \sum_{j\in \pi}w(j, {\pi}^{-1}(j))$ \\
subject to $\forall i\in \mathcal{S}$, $\pi^{-1}(i)\in \mathcal{R}_i$.
\end{minipage}
}
\end{center}
\vspace{0.1in}

Intuitively, we are looking for a feasible assortment of sponsored products that has the minimum weight. This problem can be cast  as the classical Minimum Weighted Perfect Matching problem \citep{schrijver2003combinatorial}, which can be represented as follows:
Given two sets $\mathcal{S}$ and $\mathcal{R}$,  a (not necessarily disjoint) partition of $\mathcal{R}$ into subsets $\{\mathcal{R}_1, \cdots, \mathcal{R}_{|\mathcal{S}|}\}$, and a weight function $w(i, t)$ for pairs of products $i\in \mathcal{S}$ and $t\in \mathcal{R}_i$, the goal is to find a perfect matching $\pi$ (a bijection from $\mathcal{S}$ to $\mathcal{R}$) that minimizes the total weight of matched pairs. As a result, we can employ any existing solution for the Minimum Weighted Perfect Matching problem to optimally solve $\textbf{P.4}$ and find an optimal solution $\pi^{p4}$.

Step 2: Next we compute a feasible assortment of organic products on top of the previously selected sponsored products $\pi^{p4}$. Let $w'_0=w_0+\sum_{j\in \pi^{p4}}w(j, {\pi^{p4}}^{-1}(j))$. We introduce problem $\textbf{P.5}$ as follows:

\begin{center}
\framebox[0.6\textwidth][c]{
\enspace
\begin{minipage}[t]{0.6\textwidth}
\small
$\textbf{P.5}$
\[\max_{\pi: \pi \subseteq \mathcal{O}}  \sum_{i\in \pi} \frac{r_i\cdot  w(i, \pi^{-1}(i))}{w'_0+\sum_{j\in \pi}w(j, \pi^{-1}(j)) }\]
subject to $\pi \in \mathcal{I}$.
\end{minipage}
}
\end{center}
\vspace{0.1in}

The objective of $\textbf{P.5}$ is to compute the best assortment of organic products on top of $\pi^{p4}$. After solving  $\textbf{P.5}$ approximately and obtain a solution $\pi^{p5}$, we build the final solution, denoted by $\pi^{II}$, by integrating  $\pi^{p4}$ and $\pi^{p5}$, that is, in the final solution $\pi^{II}$, we follow $\pi^{p4}$ for the allocation of sponsored products and follow $\pi^{p5}$ for the allocation of organic products.

\begin{lemma}
\label{lem:3}
Suppose $\pi^{p5}$ is a $\beta$-approximate solution for $\textbf{P.5}$, and $\pi^{II}$ is the final solution that integrates  $\pi^{p4}$ and $\pi^{p5}$, we have
\begin{eqnarray}
f(\pi^{II}) \geq \beta \cdot \sum_{i\in \pi^*\cap  \mathcal{O} } \frac{r_i\cdot  w(i, {\pi^*}^{-1}(i))}{w_0+\sum_{j\in \pi^*}w(j, {\pi^*}^{-1}(j)) }.
\end{eqnarray} I.e., $f(\pi^{II}) \geq \beta\cdot \mbox{ \texttt{part II} in  (\ref{eq:8})}$.
\end{lemma}
\emph{Proof:} Because $\pi^{II}$ combines $\pi^{p4}$ and $\pi^{p5}$, $f(\pi^{II})$ is at least as high as the utility contributed by $\pi^{p5}$. That is,
\begin{eqnarray}
\label{eq:9}
f(\pi^{II}) \geq \sum_{i\in \pi^{p5}} \frac{r_i\cdot  w(i, {\pi^{p5}}^{-1}(i))}{w'_0+\sum_{j\in \pi^{p5}}w(j, {\pi^{p5}}^{-1}(j)) }.
\end{eqnarray}

Observe that $\pi^*\cap  \mathcal{O} $ must be a feasible solution of $\textbf{P.5}$. This, together with the assumption that $\pi^{p5}$ is a $\beta$-approximate solution for $\textbf{P.5}$, implies that
\begin{eqnarray}
\sum_{i\in \pi^{p5}} \frac{r_i\cdot  w(i, {\pi^{p5}}^{-1}(i))}{w'_0+\sum_{j\in \pi^{p5}}w(j, {\pi^{p5}}^{-1}(j)) }\geq \beta \cdot  \sum_{i\in \pi^*\cap  \mathcal{O} } \frac{r_i\cdot  w(i, {\pi^*}^{-1}(i))}{w'_0+\sum_{j\in \pi^*}w(j, {\pi^*}^{-1}(j)) }.
\end{eqnarray}

This, together with inequality (\ref{eq:9}), implies that $f(\pi^{II}) \geq \beta \cdot \sum_{i\in \pi^*\cap  \mathcal{O} } \frac{r_i\cdot  w(i, {\pi^*}^{-1}(i))}{w_0+\sum_{j\in \pi^*}w(j, {\pi^*}^{-1}(j)) }$. $\Box$

\textbf{Putting it all together.} Recall from Lemma \ref{lem:2} that we demonstrate $f(\pi^I) \geq \mbox{\texttt{ part I} in  (\ref{eq:8})}$, and in Lemma \ref{lem:3}, we establish $f(\pi^{II}) \geq \beta\cdot \mbox{ \texttt{part II} in  (\ref{eq:8})}$, where $\beta$ represents the approximation ratio in solving $\textbf{P.5}$. This, together with equality (\ref{eq:8}), implies that
\begin{eqnarray}
\label{eq:10}
f(\pi^*) =  \mbox{ \texttt{part I }}+ \mbox{\texttt{ part II}}\leq f(\pi^I) + f(\pi^{II})/\beta.
\end{eqnarray}

Suppose we randomly pick a solution from $\pi^I$ and $\pi^{II}$ such that $\pi^I$ is selected with probability $1-\frac{1}{\beta+1}$ and $\pi^{II}$ is selected with probability $\frac{1}{\beta+1}$, then the expected utility of the selected solution is

\begin{eqnarray}
(1-\frac{1}{\beta+1})\cdot f(\pi^I) + \frac{1}{\beta+1} \cdot f(\pi^{II})&& = \frac{\beta}{\beta+1}\cdot f(\pi^I) + \frac{1}{\beta+1} \cdot f(\pi^{II})\\
&& =\frac{\beta}{\beta+1}\cdot (f(\pi^I) + f(\pi^{II})/\beta)\\
&&\geq \frac{\beta}{\beta+1}\cdot f(\pi^*)
\end{eqnarray}
where the inequality by inequality (\ref{eq:10}). It follows that
\begin{eqnarray}
\max\{f(\pi^I),  f(\pi^{II})\} \geq (1-\frac{1}{\beta+1})\cdot f(\pi^I) + \frac{1}{\beta+1} \cdot f(\pi^{II}) \geq \frac{\beta}{\beta+1}\cdot f(\pi^*).
\end{eqnarray}

That is, the better solution between $\pi^I$ and $\pi^{II}$ achieves an approximation ratio of $\frac{\beta}{\beta+1}$, establishing the following main result.
\begin{theorem}
If a $\beta$-approximate solution for $\textbf{P.5}$ exists, we can attain an approximation ratio of $\frac{\beta}{\beta+1}$ for the original problem $\textbf{P.2}$.
\end{theorem}

The next section is dedicated to finding an efficient solution for $\textbf{P.5}$ subject to practical constraints. As it will become clear later, if $\mathcal{I}$ contains subsets of $\mathcal{O}$ that satisfy capacity (or knapsack) constraints, we achieve an approximation ration of $\beta=1/(1+\epsilon)3$; if $\mathcal{I}$ contains subsets of $\mathcal{O}$ that satisfy partition matroid constraints, we achieve an approximation ratio of $\beta=1/(2+\epsilon)$.

\subsection{Approximation algorithm for $\textbf{P.5}$}

Recall that  $w'_0=w_0+\sum_{j\in \pi^{p4}}w(j, {\pi^{p4}}^{-1}(j))$, let $f'(\pi)=\sum_{i\in \pi} \frac{r_i\cdot  w(i, \pi^{-1}(i))}{w'_0+\sum_{j\in \pi}w(j, \pi^{-1}(j)) }$. $\textbf{P.5}$ can be written as follows:
\begin{center}
\framebox[0.4\textwidth][c]{
\enspace
\begin{minipage}[t]{0.4\textwidth}
\small
$\textbf{P.5}$
$\max_{\pi: \pi \subseteq \mathcal{O}} f'(\pi)$
subject to $\pi \in \mathcal{I}$.
\end{minipage}
}
\end{center}
\vspace{0.1in}

To solve $\textbf{P.5}$, we convert it from the original sequencing problem to a subset selection problem. To this end, we introduce a set function that operates on a ground set denoted by $\mathcal{U} = \{(i,t)\mid i\in \mathcal{O}, t\in \mathcal{P}\}$. In this context, the selection of an element $(i,t)\in \mathcal{U} $ corresponds to placing product $i$ in position $t$ for assortment planning purposes. Given a set $U\subseteq \mathcal{U}$, we define $\mathcal{O}(U)$ as the set containing all products $i\in \mathcal{I}$ for which there exists at least one element $(i,t)$ in $U$ for some $t\in \mathcal{P}$. This set is represented by $\mathcal{O}(U)=\{i\in \mathcal{O} \mid \exists t\in \mathcal{P}, (i,t) \in U\}$. For each product $i\in \mathcal{O}(U)$, we define $\omega(U, i)$ as the maximum weight among all positions $t\in \mathcal{P}$ such that $(i,t)\in U$, and it is denoted as $\omega(U, i)=\max_{ t\in \mathcal{P}: (i,t)\in U} w(i,t)$. Then we define the utility function $l: 2^\mathcal{U}\rightarrow \mathbb{R}_{\geq 0}$ as follows
\begin{eqnarray}
l(U)=\sum_{i\in \mathcal{O}(U)}\frac{r_i\cdot \omega(U, i)}{w'_0+\sum_{j\in \mathcal{O}(U)}\omega(U, j)}.
\end{eqnarray}
Intuitively, $l(U)$ captures the expected utility of an assortment specified by $U$.

We define $h(U)$ as the largest utility that can be obtained by selecting a subset of elements from $U$. That is,
\begin{eqnarray}
h(U)=\max_{X\subseteq U}l(X).
\end{eqnarray}

Let $\pi'$ denote the optimal solution of $\textbf{P.5}$ and $r'_{\min}$ represent the minimum revenue among all products in $\pi'$, formally defined as $r'_{\min} = \min_{i \in \pi'} r_i$. Define $\mathcal{U}'$ as the set of elements whose revenue is at least $r'_{\min}$, that is, $\mathcal{U}'=\{(i,t)\in\mathcal{U} \mid r_i \geq r'_{\min}\}$. Let $ \mathcal{U}'_t$ be the subset of $ \mathcal{U}' $ containing all elements that correspond to position $t$. Formally, $\mathcal{U}'_t = \{ (i, t) \in \mathcal{U}' \,|\, i \in \mathcal{O} \}$ for each $t \in \mathcal{P}$.
Next we introduce a new optimization problem over $\mathcal{U}'$. The objective of $\textbf{P.6}$ is to find a set $U\subseteq \mathcal{U}'$ that maximizes $ \min\{h(U), r'_{\min}\}$ subject to $\forall t\in \mathcal{P}$, $|U\cap \mathcal{U}'_t|\leq 1$, and
$\mathcal{O}(U)  \in \mathcal{I}$. The idea of introducing this surrogate utility function draws inspiration from \citep{el2023joint}.

 \begin{center}
\framebox[0.4\textwidth][c]{
\enspace
\begin{minipage}[t]{0.4\textwidth}
\small
$\textbf{P.6}$
$\max_{U\subseteq \mathcal{U}'}  \min\{h(U), r'_{\min}\}$ \\
subject to \\
$\forall t\in \mathcal{P}$, $|U\cap \mathcal{U}'_t|\leq 1$.\\
$\mathcal{O}(U)  \in \mathcal{I}$.
\end{minipage}
}
\end{center}
\vspace{0.1in}

Let $U^*$ denote the optimal solution of $\textbf{P.6}$, we first present an upper bound of the optimal solution of  $\textbf{P.5}$.
\begin{lemma}
\label{eq:2}
$\min\{h(U^*), r'_{\min}\} \geq f'(\pi')$
\end{lemma}
\emph{Proof:} Because $\pi'$ is the optimal solution of  $\textbf{P.5}$ and $r'_{\min}$ is the minimum revenue among all products in $\pi'$, according to Lemma \ref{lem:1} (in appendix), we have
\begin{eqnarray}f'(\pi')\leq r'_{\min}.
\end{eqnarray} Next we consider a set of elements $U'$ such that $U'=\{(i,t)\in \mathcal{U}'\mid i\in \pi' \mbox{ and } \pi'^{-1}(i)=t\}$. That is, $U'$ contains all elements that are corresponding to the assortment defined by $\pi'$. Because $\pi'$ is a feasible solution of  $\textbf{P.5}$, we have
\begin{itemize}
\item $\pi'$ places at most one product in each position, hence, we have $\forall t\in \mathcal{P}$, $|U'\cap \mathcal{U}'_t|\leq 1$;
\item $\pi' \in \mathcal{I} $, this implies that $\mathcal{O}(U')  \in \mathcal{I}$.
\end{itemize}

It follows that $U'$ is a feasible solution of $\textbf{P.6}$. Moreover,
\begin{eqnarray}
&&l(U')=\sum_{i\in \mathcal{O}(U')}\frac{r_i\cdot \omega(U', i)}{w'_0+\sum_{i\in \mathcal{O}(U')}w(j, \pi'^{-1}(i))}\\
&&=\sum_{i\in \mathcal{O}(U')}\frac{r_i\cdot w(i, \pi'^{-1}(i))}{w'_0+\sum_{j\in \mathcal{O}(U')}w(j, \pi'^{-1}(j))}\\
&&= f'(\pi'),
\end{eqnarray}
where the second equality is by the observation that $\forall t\in \mathcal{P}$, $|U'\cap \mathcal{U}'_t|\leq 1$ and the last equality is by the definition of $f'(\pi')$. This indicates that
\begin{eqnarray}
h(U') \geq l(U')= f'(\pi')
\end{eqnarray}
where the inequality is by the definition of $h$, which states that $h(U')=\max_{X\subseteq U'}l(X)$.

This, together with inequality (\ref{eq:2}), implies that
\begin{eqnarray}
\label{eq:3}
\min\{h(U'), r'_{\min}\} \geq f'(\pi').
\end{eqnarray}

Because  $U'$ is a feasible solution of $\textbf{P.6}$ and $U^*$ is the optimal solution of  $\textbf{P.6}$, we have
\begin{eqnarray}
\min\{h(U^*), r'_{\min}\} \geq \min\{h(U'), r'_{\min}\} \geq f'(\pi')
\end{eqnarray}
where the second inequality is by inequality (\ref{eq:3}). $\Box$

Next, we demonstrate that the objective function $\min\{h(U), r'_{\min}\}$ of $\textbf{P.6}$, defined over elements from $\mathcal{U}'$, exhibits monotonicity and submodularity\footnote{A function $f: 2^V \rightarrow \mathbb{R}$ is a submodular function if, for all $A, B \subseteq V$ where $A \subseteq B$, and for all $x \in V \setminus B$, the following condition holds: $f(A \cup \{x\}) - f(A) \geq f(B \cup \{x\}) - f(B)$.

}. To establish this property, we can alternatively prove the following lemma.

\begin{lemma}
 Let $r_{\min}$ represent the minimum revenue among all organic products, with $r_{\min}=\min_{i\in \mathcal{O}} r_i$, The set function $U\rightarrow \min\{h(U), r_{\min}\}$ is a monotone and submodular function on $\mathcal{U}$.
\end{lemma}
\emph{Proof:} Consider two sets $X, Y \subseteq \mathcal{U}$ such that $X \subseteq Y$, and let $(i,t) \notin Y$. The rest of the proof is devoted to proving the following
inequality
\begin{eqnarray}
g((i,t)\mid Y) \leq g((i,t)\mid X).
\end{eqnarray}
We proceed by considering three cases, following a similar approach to the proof of Lemma 4.3 in \citep{el2023joint}. However, additional care is required since our function $h$ is defined differently from theirs.

Case 1: Suppose that $h(Y)\geq r_{\min}$. In this case, by the monotonicity of $h$, we have $h(Y+(i,t))\geq h(Y) \geq r_{\min}$. It follows that $g(Y+(i,t))= g(Y) = r_{\min}$. Therefore, we have $g((i,t)\mid Y)=0$. It follows that $g((i,t)\mid Y)=0\leq g((i,t)\mid X)$ where the inequality is by the fact that $g$ is non-decreasing.

Case 2: Suppose that $h(Y+(i,t))\geq r_{\min}\geq h(Y)$. In this case, we have $g((i,t)\mid Y)=   r_{\min}- h(Y)$. Because $h$ is non-decreasing, we have $h(X)\leq  h(Y) \leq r_{\min}$. Hence, $g(X)=h(X)$. If we have $g(X+(i,t))= r_{\min}$, then $g((i,t)\mid X)=   r_{\min}- h(X)$. Because $ h(Y)\geq h(X)$ by the monotonicity of $h$, we have $g((i,t)\mid Y) \leq g((i,t)\mid X)$. In the rest of this case, we assume that $g(X+(i,t))= h(X+(i,t))$.

Provided that $g(X+(i,t))= h(X+(i,t))$, we have $ h(X+(i,t))\leq r_{\min}$, and hence, $ h(X)\leq r_{\min}$  by the monotonicity of $h$. Because $r_{\min}$ is the smallest revenue among all organic products, we must have $h(X+(i,t))= l(X+(i,t))$ and $h(X)= l(X)$. Similarly, because $h(Y) \leq r_{\min}$, we have $h(Y)=l(Y)$. Let $\Delta=\omega(X+(i,t), i) - \omega(X, i)$ and $\Delta'=\omega(Y+(i,t), i) - \omega(Y, i)$, we have 
\begin{eqnarray}\label{eq:lake222}
\Delta \geq \Delta'
\end{eqnarray}. This is because $\omega(U, i)=\max_{ t\in \mathcal{P}: (i,t)\in U} w(i,t)$ is defined as the maximum weight over all positions in $U$  for a given $i$. Therefore, it is a monotone and submodular function for all $i$. Since $X \subseteq Y$, inequality (\ref{eq:lake222}) follows.

Observe that
\begin{eqnarray}
&&g((i,t)\mid X) = l(X+(i,t))-l(X)\\
&& =\frac{\sum_{j\in \mathcal{O}(X+(i,t))} r_j\cdot \omega(X+(i,t), j)}{w'_0+\sum_{j\in \mathcal{O}(X+(i,t))}\omega(X+(i,t), j)}- \frac{\sum_{j\in \mathcal{O}(X)} r_j\cdot \omega(X, j)}{w'_0+\sum_{j\in \mathcal{O}(X)}\omega(X, j)}\\
&&= \frac{r_i\cdot \Delta+\sum_{j\in \mathcal{O}(X)} r_j\cdot \omega(X, j)}{w'_0+\Delta + \sum_{j\in \mathcal{O}(X)}\omega(X, j)}- \frac{\sum_{j\in \mathcal{O}(X)} r_j\cdot \omega(X, j)}{w'_0+\sum_{j\in \mathcal{O}(X)}\omega(X, j)}\\
&&= \frac{\Delta}{w'_0+\Delta + \sum_{j\in \mathcal{O}(X)}\omega(X, j)}\cdot(r_i- l(X))\\
&&\geq \frac{\Delta'}{w'_0+\Delta' + \sum_{j\in \mathcal{O}(X)}\omega(X, j)}\cdot(r_i- l(X))\\
&&\geq \frac{\Delta'}{w'_0+\Delta' + \sum_{j\in \mathcal{O}(Y)}\omega(Y, j)}\cdot(r_i- l(Y))\\
&&= \frac{r_i\cdot \Delta'+\sum_{j\in \mathcal{O}(Y)} r_j\cdot \omega(Y, j)}{w'_0+\Delta' + \sum_{j\in \mathcal{O}(Y)}\omega(Y, j)}- \frac{\sum_{j\in \mathcal{O}(Y)} r_j\cdot \omega(Y, j)}{w'_0+\sum_{j\in \mathcal{O}(Y)}\omega(Y, j)}\\
&&= \frac{r_i\cdot \Delta'+\sum_{j\in \mathcal{O}(Y)} r_j\cdot \omega(Y, j)}{w'_0+\Delta' + \sum_{j\in \mathcal{O}(Y)}\omega(Y, j)}- h(Y)\\
&&= l(Y+(i,t))- h(Y). \label{eq:lake}
\end{eqnarray}

The first inequality is by the fact that $\Delta \geq \Delta'$ (inequality (\ref{eq:lake222})). 

Now we have \begin{eqnarray}
&&g((i,t)\mid X) \geq l(Y+(i,t))- h(Y). \label{eq:lake4}
\end{eqnarray} To prove this case, it is sufficient to prove that $l(Y+(i,t))\geq r_{\min}$.  This is because we have $g((i,t)\mid Y)=   r_{\min}- h(Y)$, if $l(Y+(i,t))\geq r_{\min}$ holds, then inequality (\ref{eq:lake4}) implies that $g((i,t)\mid X)\geq l(Y+(i,t))- h(Y) \geq r_{\min}- h(Y)= g((i,t)\mid Y)$. We next show that $l(Y+(i,t))\geq r_{\min}$. The proof follows the same reasoning as in the proof of Lemma 4.3 in \citep{el2023joint}. Recall that we assume  $h(Y+(i,t))\geq r_{\min}$, this implies that
\begin{eqnarray}
h(Y+(i,t))=\max_{Z \subseteq Y+(i,t)}l(Z)\geq r_{\min}.
\end{eqnarray}
Let $Z^*=\argmax_{Z \subseteq Y+(i,t)}l(Z)$, we have
\begin{eqnarray}
&&l(Y+(i,t))=  \sum_{j\in \mathcal{O}(Y+(i,t))}\frac{r_j\cdot \omega(Y+(i,t), j)}{w'_0+\sum_{j\in \mathcal{O}(Y+(i,t))}\omega(Y+(i,t), j)}\\
&& =  \frac{\sum_{j\in \mathcal{O}(Z^*)} r_j\cdot \omega(Y+(i,t), j)+ \sum_{j\in \mathcal{O}(Y+(i,t)\setminus Z^*)} r_j\cdot \omega(Y+(i,t), j)}{w'_0+\sum_{j\in \mathcal{O}(Z^*)} \omega(Y+(i,t), j)+ \sum_{j\in \mathcal{O}(Y+(i,t)\setminus Z^*)}  \omega(Y+(i,t), j)}\\
&&= \frac{\sum_{j\in \mathcal{O}(Z^*)} r_j\cdot \omega(Z^*, j)+ \sum_{j\in \mathcal{O}(Y+(i,t)\setminus Z^*)} r_j\cdot \omega(Y+(i,t)\setminus Z^*, j)}{w'_0+\sum_{j\in \mathcal{O}(Z^*)} \omega(Z^*, j)+ \sum_{j\in \mathcal{O}(Y+(i,t)\setminus Z^*)}  \omega(Y+(i,t)\setminus Z^*, j)}. \label{essay1-eq:1}
\end{eqnarray}

Because $l(Z^*)\geq r_{\min}$, we have $\frac{\sum_{j\in \mathcal{O}(Z^*)} r_j\cdot \omega(Z^*, j)}{w'_0+\sum_{j\in \mathcal{O}(Z^*)} \omega(Z^*, j)}\geq r_{\min}$. Moreover, we have $\forall j\in \mathcal{O}(Y+(i,t)\setminus Z^*), r_j \geq  r_{\min}$. This, together with equality (\ref{essay1-eq:1}), implies that
\begin{eqnarray}
&&l(Y+(i,t))= \\
&&\frac{\sum_{j\in \mathcal{O}(Z^*)} r_j\cdot \omega(Z^*, j)+ \sum_{j\in \mathcal{O}(Y+(i,t)\setminus Z^*)} r_j\cdot \omega(Y+(i,t)\setminus Z^*, j)}{w'_0+\sum_{j\in \mathcal{O}(Z^*)} \omega(Z^*, j)+ \sum_{j\in \mathcal{O}(Y+(i,t)\setminus Z^*)}  \omega(Y+(i,t)\setminus Z^*, j)} \\
&&\geq r_{\min}.
\end{eqnarray}

Case 3.  Suppose that $r_{\min} \geq h(Y+(i,t)) \geq h(Y)$. In this case, we have $g((i,t)\mid Y)=  h(Y+(i,t))- h(Y) = l(Y+(i,t))- l(Y)$ where the second equality is by the following observation: because we assume that both $h(Y+(i,t))$ and $h(Y)$ are no greater than $r_{\min}$, which is the smallest revenue, we have $\forall (i',t')\in Y+(i,t), r_{i'}\geq h(Y+(i,t)) \mbox{ and } r_{i'}\geq  h(Y)$. Then according to Lemma \ref{lem:1} (in appendix), we conclude that $h(Y+(i,t))=l(Y+(i,t))$ and $h(Y)=l(Y)$. Because $h$ is non-decreasing, we have $h(X)\leq  h(Y) \leq r_{\min}$ and $h(X+(i,t))\leq  h(Y+(i,t)) \leq r_{\min}$. Thus, $g((i,t)\mid X)=  h(X+(i,t))- h(X) = l(X+(i,t))- l(X)$ where the second equality can be proved using the previous argument, that is, we can show that $h(X+(i,t))=l(X+(i,t))$ and $h(X)=l(X)$.

Let $\Delta=\omega(X+(i,t), i) - \omega(X, i)$ and $\Delta'=\omega(Y+(i,t), i) - \omega(Y, i)$, it follows that
\begin{eqnarray}
&&l(X+(i,t))-l(X)\\
&& =\frac{\sum_{j\in \mathcal{O}(X+(i,t))} r_j\cdot \omega(X+(i,t), j)}{w'_0+\sum_{j\in \mathcal{O}(X+(i,t))}\omega(X+(i,t), j)}- \frac{\sum_{j\in \mathcal{O}(X)} r_j\cdot \omega(X, j)}{w'_0+\sum_{j\in \mathcal{O}(X)}\omega(X, j)}\\
&&= \frac{r_i\cdot \Delta+\sum_{j\in \mathcal{O}(X)} r_j\cdot \omega(X, j)}{w'_0+\Delta + \sum_{j\in \mathcal{O}(X)}\omega(X, j)}- \frac{\sum_{j\in \mathcal{O}(X)} r_j\cdot \omega(X, j)}{w'_0+\sum_{j\in \mathcal{O}(X)}\omega(X, j)}\\
&&= \frac{\Delta}{w'_0+\Delta + \sum_{j\in \mathcal{O}(X)}\omega(X, j)}\cdot(r_i- l(X))\\
&&\geq \frac{\Delta'}{w'_0+\Delta' + \sum_{j\in \mathcal{O}(X)}\omega(X, j)}\cdot(r_i- l(X))\\
&&\geq \frac{\Delta'}{w'_0+\Delta' + \sum_{j\in \mathcal{O}(Y)}\omega(Y, j)}\cdot(r_i- l(Y))\\
&&= l(Y+(i,t))- l(Y).
\end{eqnarray} The first inequality is by the fact that $\Delta \geq \Delta'$ (inequality (\ref{eq:lake222})). $\Box$

Now we are in position to present the main theorem of this section.
\begin{theorem}
\label{thm:2}
If there exists an $\alpha$-approximation algorithm for problem $\textbf{P.6}$, then there exists an $\alpha$-approximation algorithm for problem $\textbf{P.5}$.
\end{theorem}
\emph{Proof:} Let $U^*$ denote the optimal solution of $\textbf{P.6}$. Assume there exists a solution $U\subseteq \mathcal{U}'$ such that $U$ is a feasible solution of  $\textbf{P.6}$ and $\min\{h(U), r'_{\min}\}\geq \alpha \min\{h(U^*), r'_{\min}\}$. We first show that
\begin{eqnarray}
\label{eq:5}
h(U)\leq r'_{\min}.
 \end{eqnarray} Let $V=\argmax_{X\subseteq U} l(X)$. By the definition of $h(U)$, we have $h(U)=l(V)$. Because  $V$ is a subset of $U$ and the constraint of problem $\textbf{P.6}$ is downward-closed, $V$ must be a feasible solution of $\textbf{P.6}$. We next construct an assortment planning $\pi$ over $V$ such that for each product $i\in V$, set $\pi^{-1}(i)=\argmax_{t: (i,t)\in V} w(i,t)$. That is, $\pi$ places $i\in V$ in the best position specified in $V$. By the definition of $l(V)$, we have
\begin{eqnarray}\label{eq:4}
l(V)=f'(\pi).
 \end{eqnarray} Moreover, $\pi$ is a feasible solution of $\textbf{P.5}$. By the assumption that $\pi'$ is an optimal solution of $\textbf{P.5}$, we have $f'(\pi')\geq f'(\pi)$. This, together with inequality (\ref{eq:4}), implies that  $f'(\pi')\geq f'(\pi)=l(V)$. Because $f'(\pi')\leq  r'_{\min}$, we have $l(V)\leq f'(\pi')\leq r'_{\min}$. It follows that $h(U)=l(V)\leq r'_{\min}$. This finishes the proof of inequality (\ref{eq:5}).

 Given inequality (\ref{eq:5}) and the assumption that  $\min\{h(U), r'_{\min}\}\geq \alpha \min\{h(U^*), r'_{\min}\}$, we have $\min\{h(U), r'_{\min}\} = h(U)\geq \alpha \min\{h(U^*), r'_{\min}\}$. This, together with Lemma \ref{eq:2}, implies that
 \begin{eqnarray}
 \label{eq:7}
 h(U)\geq \alpha \min\{h(U^*), r'_{\min}\}\geq \alpha f'(\pi').
 \end{eqnarray}

 Now we are ready to present an algorithm that finds  a solution for problem $\textbf{P.5}$ with a utility value at least $\alpha f'(\pi')$. We first apply an $\alpha$-approximation algorithm to solve problem $\textbf{P.6}$ to obtain a solution $U$. Then we compute $V=\argmax_{X\subseteq U} l(X)$, which can be done in polynomial time \citep{talluri2004revenue}. At last, we compute an assortment planning $\pi$ over $V$ such that for each product $i\in V$, set $\pi^{-1}(i)=\argmax_{t: (i,t)\in V} w(i,t)$. That is, $\pi$ places $i\in V$ in the best position specified in $V$. Inequalities (\ref{eq:4}) and (\ref{eq:7}) together imply that $f'(\pi)=l(V)=h(U)\geq \alpha f'(\pi')$ where the second equality is by the definition of $h(U)$. $\Box$

\textbf{Discussion on two example constraints.} We next discuss two important examples of $\mathcal{I}$.  One example of $\mathcal{I}$ contains subsets of $\mathcal{O}$ that satisfy capacity (or knapsack) constraints. For example, assuming each product $i\in \mathcal{O}$ has a cost of $c_i$, and we are also given a capacity constraint $C$. In this context, the feasible family
 $\mathcal{I}$ can be defined as  $\mathcal{I}=\{A\subseteq \mathcal{O} \mid \sum_{i\in A} c_i \leq C\}$. Then  $\textbf{P.6}$ can be written as
 \begin{center}
\framebox[0.4\textwidth][c]{
\enspace
\begin{minipage}[t]{0.4\textwidth}
\small
$\max_{U\subseteq \mathcal{U}'}  \min\{h(U), r'_{\min}\}$ \\
subject to \\
$\forall t\in \mathcal{P}$, $|U\cap \mathcal{U}'_t|\leq 1$ and
$\sum_{i\in \mathcal{O}(U) } c_i \leq C$.
\end{minipage}
}
\end{center}
\vspace{0.1in}

We can covert this problem to the following problem.
 \begin{center}
\framebox[0.4\textwidth][c]{
\enspace
\begin{minipage}[t]{0.4\textwidth}
\small
$\max_{U\subseteq \mathcal{U}'}  \min\{h(U), r'_{\min}\}$ \\
subject to \\
$\forall t\in \mathcal{P}$, $|U\cap \mathcal{U}'_t|\leq 1$ and
$\sum_{(i,t)\in U} c_i \leq C$.
\end{minipage}
}
\end{center}
\vspace{0.1in}

Because its objective function is monotone and submodular, the above problem is to maximize a monotone submodular function subject to a partition matroid (e.g., $\forall t\in \mathcal{P}$, $|U\cap \mathcal{U}'_t|\leq 1$) constraint and a knapsack constraint (e.g., $\sum_{(i,t)\in U} c_i \leq C$). The state-of-the-art algorithm \cite{10.1609/aaai.v37i4.25510} achieves an approximation ratio of $1/(1+\epsilon)3$. Hence, we have $\alpha=1/(1+\epsilon)3$ in Theorem \ref{thm:2} for this case.

 The second example of $\mathcal{I}$ contains subsets of $\mathcal{O}$ that satisfy partition matroid constraints. For example, assuming all organic products are partitioned into $m$ groups $\{\mathcal{G}_1, \mathcal{G}_2, \cdots, \mathcal{G}_m\}$. The feasible family
 $\mathcal{I}$ can be defined as  $\mathcal{I}=\{A\subseteq \mathcal{O} \mid \forall q\in[m], |A\cap \mathcal{G}_q|\leq \gamma_q\}$, where $\gamma_q$ represents an upper limit on the number of products that can be chosen from group $\mathcal{G}_q$. This formulation is useful for incorporating additional criteria into product selection, such as considerations of fairness and diversity. In this case,  $\textbf{P.6}$ can be written as
 \begin{center}
\framebox[0.6\textwidth][c]{
\enspace
\begin{minipage}[t]{0.6\textwidth}
\small
$\max_{U\subseteq \mathcal{U}'}  \min\{h(U), r'_{\min}\}$ \\
subject to \\
$\forall t\in \mathcal{P}$, $|U\cap \mathcal{U}'_t|\leq 1$ and
$\forall q\in[m], |\mathcal{O}(U) \cap \mathcal{G}_q|\leq \gamma_q$.
\end{minipage}
}
\end{center}
\vspace{0.1in}

Let  $\mathcal{U}_q=\{(i,t)\mid i\in \mathcal{G}_q\}$ for all $q\in[m]$, we can covert the above problem to the following problem.
 \begin{center}
\framebox[0.6\textwidth][c]{
\enspace
\begin{minipage}[t]{0.6\textwidth}
\small
$\max_{U\subseteq \mathcal{U}'}  \min\{h(U), r'_{\min}\}$ \\
subject to \\
$\forall t\in \mathcal{P}$, $|U\cap \mathcal{U}'_t|\leq 1$ and
$\forall q\in[m], |U \cap \mathcal{U}_q|\leq \gamma_q$.
\end{minipage}
}
\end{center}
\vspace{0.1in}

The above problem is to maximize a monotone submodular function subject to two partition matroid constraints (e.g., $\forall t\in \mathcal{P}$, $|U\cap \mathcal{U}'_t|\leq 1$ and $\forall q\in[m], |U \cap \mathcal{U}_q|\leq \gamma_q$). The state-of-the-art algorithm \citep{lee2010submodular}  gives an approximation ratio of $1/(2 +\epsilon)$. Hence, we have $\alpha=1/(2 +\epsilon)$ in Theorem \ref{thm:2} for this case.

\section{Conclusion}
In conclusion, this research aims to address the critical challenge of assortment planning in the presence of sponsored products. We contribute to the development of efficient and effective assortment planning that maximize the platform revenue, while ensuring the fulfillment of advertiser requirements. Our findings pave the way for future research in this domain and offer valuable guidelines for businesses seeking to navigate the complexities of sponsored product integration in their e-commerce platforms and recommendation systems.

\bibliographystyle{ijocv081}
\bibliography{reference}

\section{Appendix}
The following lemma is a classical result in standard assortment optimization \cite{el2023joint, talluri2004revenue,gallego2004managing}. We show that, with slight modifications to the proof, specifically to account for the function $\omega$, the result also extends to our setting.

\begin{lemma}
\label{lem:1}
Consider the problem $\max_{X\subseteq U}l(X)$ (labeled as $\textbf{P.7}$), let $l^*$ denote the optimal objective value of  $\textbf{P.7}$. There exists an optimal solution  $X^*$  of  $\textbf{P.7}$ such that
\begin{eqnarray}
X^*=\{(i,t)\in U \mid r_i \geq l^*\}.
\end{eqnarray}
\end{lemma}
\emph{Proof:} Consider an arbitrary optimal solution $X$ of $\textbf{P.7}$ with $l(X)=l^*$. If $X=\{(i,t)\in U \mid r_i \geq l^*\}$, then this lemma is proved. Next we prove the case if $X$ does not satisfy this condition. We show that for any such $X$, we can construct another optimal solution that satisfies the aforementioned condition.

Consider an element $(i,t)$ such that $(i,t)\notin X$ and $r_i\geq l^*$. Let $\Delta=\omega(X+(i,t), i) - \omega(X, i)$, it follows that
\begin{eqnarray}
&&l(X+(i,t))-l(X)\\
&& =\frac{\sum_{j\in \mathcal{O}(X+(i,t))} r_j\cdot \omega(X+(i,t), j)}{w'_0+\sum_{j\in \mathcal{O}(X+(i,t))}\omega(X+(i,t), j)}- \frac{\sum_{j\in \mathcal{O}(X)} r_j\cdot \omega(X, j)}{w'_0+\sum_{j\in \mathcal{O}(X)}\omega(X, j)}\\
&&= \frac{r_i\cdot \Delta+\sum_{j\in \mathcal{O}(X)} r_j\cdot \omega(X, j)}{w'_0+\Delta + \sum_{j\in \mathcal{O}(X)}\omega(X, j)}- \frac{\sum_{j\in \mathcal{O}(X)} r_j\cdot \omega(X, j)}{w'_0+\sum_{j\in \mathcal{O}(X)}\omega(X, j)}\\
&&= \frac{\Delta}{w'_0+\Delta + \sum_{j\in \mathcal{O}(X)}\omega(X, j)}\cdot(r_i- l(X))\\
&&= \frac{\Delta}{w'_0+\Delta + \sum_{j\in \mathcal{O}(X)}\omega(X, j)}\cdot(r_i- l^*) \label{eq:lake2} \\
&&\geq 0
\end{eqnarray}
where the forth equality is by the assumption that $l(X)=l^*$ and the last equality is by the assumption that $r_i\geq l^*$. This implies that the inclusion of $(i,t)$ into $X$ does not lead to a reduction in utility. By repeatedly employing this reasoning, one can readily confirm that the introduction of all $(i,t)$, where $(i,t)\notin X$ and $r_i\geq l^*$, to $X$, preserves the optimality of the solution.

On the other hand, it is easy to verify that $X$ must not contain any elements whose revenue is less than $l^*$. We prove this through contradiction. Consider an element $(i,t)$ such that $(i,t)\in X$ and $r_i< l^*$. Let $\Delta=\omega(X, j) - \omega(X\setminus (i,t), j)$. Employing a similar line of reasoning used to establish equality (\ref{eq:lake2}), we can derive the following:
\begin{eqnarray}
l(X)-l(X\setminus (i,t))= \frac{\Delta}{w'_0+\Delta + \sum_{j\in \mathcal{O}(X\setminus (i,t))}\omega(X\setminus (i,t), j)}\cdot(r_i- l^*).
\end{eqnarray}

This indicates that if $(i,t)\in X$ and $r_i< l^*$, then removing such element from $X$  leads to an increase in utility.  This contradicts to the assumption that $X$ is the optimal solution.

Now we are ready to conclude that there exists an optimal solution  $X^*$  of  $\textbf{P.7}$ such that
$
X^*=\{(i,t)\in U \mid r_i \geq l^*\}$. $\Box$



\end{document}